# Comparison of device structures for the dielectric breakdown measurement of hexagonal boron nitride

Yoshiaki Hattori,[1,a)] Takashi Taniguchi,[2)] Kenji Watanabe,[2)] and Kosuke Nagashio[1,3,b)]

[1]*Department of Materials Engineering, The University of Tokyo, Tokyo 113-8656, Japan*
[2] *National Institute of Materials Science, Ibaraki 305-0044, Japan*
[3] *PRESTO, Japan Science and Technology Agency (JST), Tokyo 113-8656, Japan*
[a)] hattori@adam.t.u-tokyo.ac.jp, [b)] nagashio@material.t.u-tokyo.ac.jp

Improving the film quality in the synthesis of large-area hexagonal boron nitride films (*h*-BN) for two-dimensional material devices remains a great challenge. The measurement of electrical breakdown dielectric strength ($E_{BD}$) is one of the most important methods to elucidate the insulating quality of *h*-BN. In this work, the $E_{BD}$ of high quality exfoliated single-crystal *h*-BN was investigated using three different electrode structures under different environmental conditions to determine the ideal electrode structure and environment for $E_{BD}$ measurement. A systematic investigation revealed that $E_{BD}$ is not sensitive to contact force or electrode area but strongly depends on the relative humidity during measurement. Once the measurement environment is properly managed, it was found that the $E_{BD}$ values are consistent within experimental error regardless of the electrode structure, which enables the evaluation of the crystallinity of synthesized *h*-BN at the microscopic and macroscopic level by utilizing the three different electrode structures properly for different purposes.

Following the many advances in the science and application of graphene and other two-dimensional (2D) materials, van der Waals (vdW) heterostructure devices formed by stacking atomically thin layered 2D crystals have attracted significant interest because of their fascinating electrical, optical and mechanical properties. The atoms connect to each other in a unit layer with strong covalent bonds, while each layer is stabilized by weak vdW forces, which enables a wide variety of stacking without the limitation of lattice mismatch at the interface.[1] Hexagonal boron nitride (*h*-BN) with its wide band gap (5.2 – 5.9 eV [2]) has been commonly used as an ideal insulating material in vdW heterostructure devices. Because insulating properties are important in electrical device applications, especially for gate insulators in field-effect transistors, fundamental research on the electrical breakdown strength ($E_{BD}$) in the out-of-plane direction has been conducted using small flakes that were mechanically exfoliated from high quality single crystals, and the $E_{BD}$ has been reported as ~10 – 12 MV/cm.[3-6] Recently, the in-plane $E_{BD}$ has also been determined to be ~3 MV/cm,[4] suggesting strong anisotropy in $E_{BD}$.

For practical applications, the large-area synthesis of *h*-BN has been intensively studied using various methods.[7-14] However, the out-of-plane $E_{BD}$ values for scalable *h*-BN are largely scattered and limited to ~4 MV/cm, which is lower than that of the exfoliated sample.[7-10] It is evident that there are common problems with impurities, and the presence of wrinkles and grain boundaries and the non-uniform thickness, which largely originate from the high nucleation density.[15] However, the out-of-plane $E_{BD}$ has been measured under different environmental conditions and by various electrode structures, such as the crossed electrode structure,[6-10] conductive atomic-force microscopy (C-AFM)[3, 5, 11] and probing systems.[12] To fairly compare the film quality, common conditions to evaluate the out-of-plane $E_{BD}$ must be maintained. In this study, the out-of-plane $E_{BD}$ is measured by different methods using small flakes that were mechanically exfoliated from high quality single crystals, since few studies as of yet have focused on measurement methods. The objective of this research is to propose an ideal electrode structure and environment for the out-of-plane $E_{BD}$ measurement.

Hereafter, the out-of-plane $E_{BD}$ is simply called $E_{BD}$. **Figure 1** compares the three different measurements and includes schematic drawings, optical microscopy images before and after the breakdown test, and experimental conditions. In all measurements, the breakdown test was conducted using the same electrical circuit with a semiconductor device parameter analyzer (B1500A, Keysight Technology). A voltage was applied to the *h*-BN and was gradually increased until breakdown with a voltage step of 50 mV and a ramping rate of 1.0 V/s. Then, the applied voltage was terminated manually after breakdown. Since the $E_{BD}$ value was not affected by the polarity of the voltage in the type B measurement, which has the highest structural symmetry, the difference in the electrode material and the work function of the metal is ignored.[16] In the actual experiment, the top electrode was grounded in all measurements. $E_{BD}$ was defined simply by $E_{BD} = V_{BD} / t_{h\text{-}BN}$, where $V_{BD}$ and $t_{h\text{-}BN}$ are the breakdown voltage and the thickness of $h$-BN, respectively.

Using the type A measurement, the breakdown test of $h$-BN on a Pt/Si substrate was conducted in contact mode of C-AFM using a rhodium-coated tip with a radius of 100 nm and a spring constant of 1.9 N/m in ambient environment (50 % relative humidity, ~20 °C). This experimental setup is the same as was used in our previous study,[3] except that the protective resistor was removed to achieve a fair comparison. At breakdown, the Pt substrate was melted locally at the AFM contact point by strong over current, and the $h$-BN formed a flower-like shape due to tearing.[3] The type A measurement allows for the control of contact force and the repetition of the breakdown test for the same $h$-BN flake.

In the type B measurement, $h$-BN was sandwiched between two electrodes. After the bottom electrode (15-nm Cr / 15-nm Au) was fabricated by electron beam (EB) lithography on a SiO$_2$/Si substrate, the $h$-BN flake was positioned on the bottom electrode using a transfer technique. Then, the top electrode (15-nm Cr / 30-nm Au) was patterned on the $h$-BN. Because the electrode pads are far from the $h$-BN flake (see Supplementary Figure S1), the breakdown test can be conducted easily using the prober in a chamber with controlled environmental conditions, such as temperature, ambient pressure, humidity and gases. As shown in the optical image after breakdown, both of the top and bottom electrodes were melted by strong Joule heating and partially peeled off. A thicker $h$-BN flake tends to cause the intense fracture of both $h$-BN and the electrode.

In the type C measurement, the top electrode (15-nm Cr / 100-nm Au) with the largest possible area was patterned by EB lithography, and then the tip of a manual prober was put in contact. It should be noted that a tungsten wire 10 μm in diameter and 2 mm in length was glued to the tip of the probe to achieve precise contact with the metal electrode on $h$-BN (Supplementary Figure S2). In this case, the probe pressure was too low to scratch the metal electrode. As shown in the optical image after breakdown, a unique electrical treeing occurs that propagates from the test point at breakdown, which is often seen in high-voltage solid discharge.[17] The pattern is formed from the thermally melted Pt substrate.

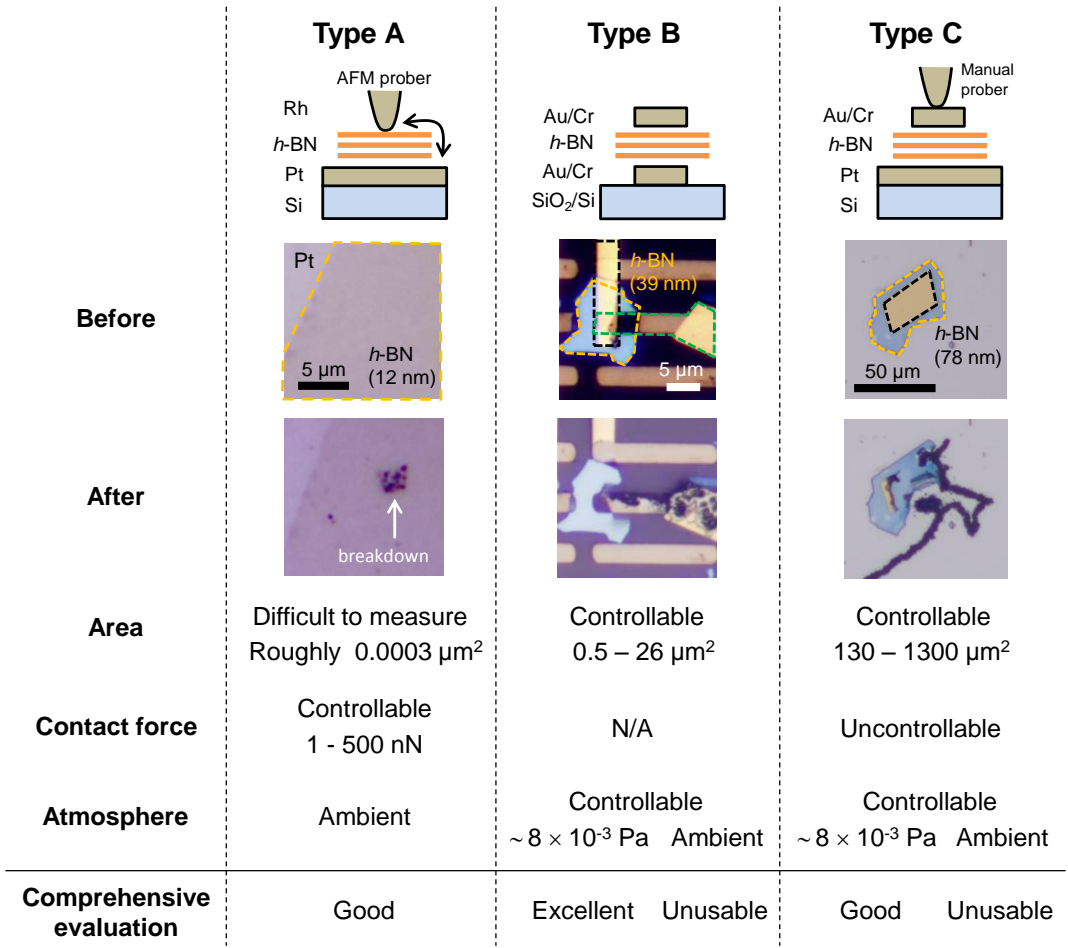

**Figure 1.** The three electrode structures used for the breakdown tests, including schematic drawings, optical microscopy images before and after the breakdown test, and the experimental conditions. The contrast in the optical images was highly emphasized by image processing. The black, yellow, and green dotted lines in the optical images indicate the top electrode, $h$-BN, and the bottom electrode, respectively. The area of the electrode is defined as the area where $h$-BN is sandwiched between the two electrodes.

Now, let us discuss the *IV* characteristics obtained by the three different methods. The typical *IV* data for ~48-nm $h$-BN is shown in **Figure 2(a)**. Tests of measurement types B and C were conducted in a vacuum of ~8 × 10$^{-3}$ Pa, while the test for measurement type A was conducted in air. The arrows in the figure indicate breakdown. The Fowler–Nordheim (FN) tunneling current before breakdown was clearly measured in the type B and C measurements, but it was not clear in the type A measurement due to the low current level, which was limited by the small contact area. Then, using the three experimental setups, the effects of contact force, ambient environment, and electrode size on the dielectric breakdown were investigated.

The contact force can be easily controlled in the type A measurement. In general, the breakdown test is conducted by C-AFM with a contact force of ~2 nN, which is similar to the value in the literature.[3, 5] After the C-AFM tip approached the $h$-BN surface, the contact force was increased by elevating the piezo stage from 1 nm to 250 nm and can be calculated by multiplying the distance of the piezo stage movement by the spring constant of the AFM tip.[18, 19] **Figure 2(b)** shows the $E_{BD}$ as a function of contact force for 13-nm $h$-BN. $E_{BD}$ is roughly independent of the contact force within experimental error. Here, the actual depth of indentation of $h$-BN was estimated by the classical Hertz model, one of the most basic contact mechanics models, which is used to find the indentation depth, contact area and contact pressure under a certain applied force when two surfaces are in contact. The values of the elastic moduli (30 GPa for $h$-BN[20] and 275 GPa for rhodium[21]), the Poisson's ratios (0.2 for $h$-BN[22] and 0.26 for rhodium[21]) and the AFM tip radius (100 nm) are used in the calculation. The actual depth of indentation for a contact force of 2 nN can be estimated to be ~0.03 nm with a contact pressure of ~0.2 GPa and a contact area of $1 \times 10^{-17}$ m$^2$. Increasing the contact force to 400 nN results an actual indentation depth of ~1 nm with a contact pressure of ~1 GPa and a contact area of $3 \times 10^{-16}$ m$^2$. Although we expect that the interlayer distance of $h$-BN is reduced by the contact force, this change was not experimentally obvious in the $E_{BD}$ value.

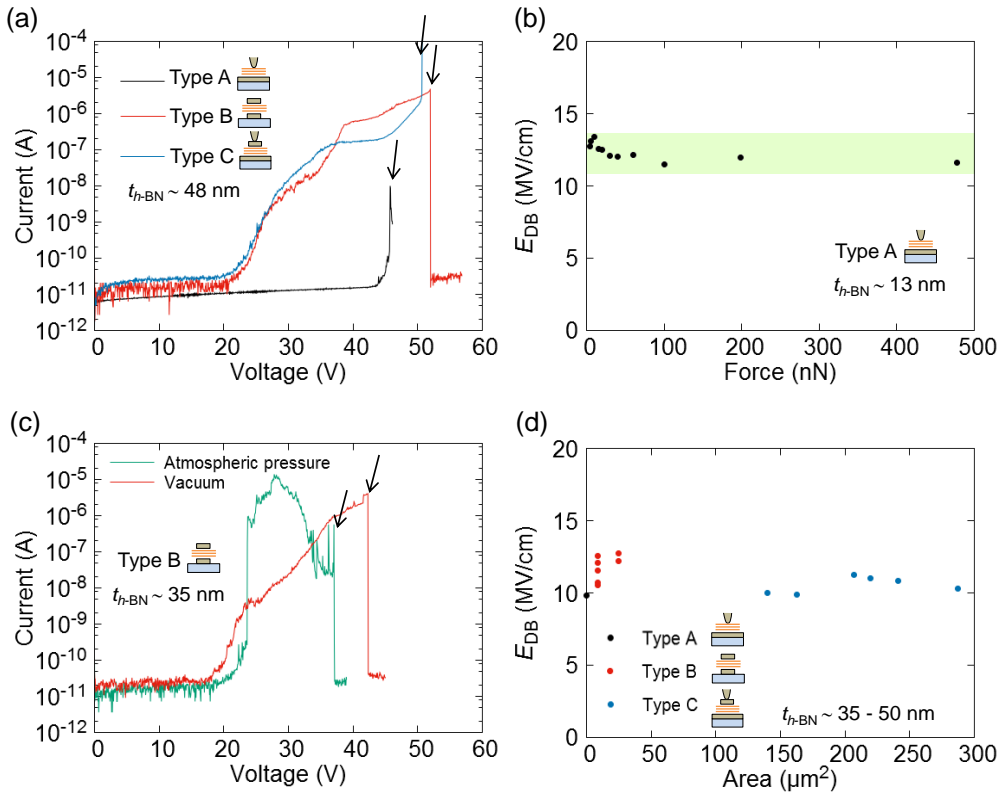

**Figure 2.** Electrical measurements. Each color has a common interpretation in all four figures, **a-d**. Black and blue represent the type A measurement and the type C measurement, respectively, while red and green represent the type B measurement under vacuum and atmospheric environment, respectively. (**a**) *IV* curves of the three methods for ~48-nm $h$-BN. The arrows in the figures indicate the breakdown voltage. Type B and C measurements were conducted under vacuum, while the type A measurement was conducted in air. In the type B measurement, the current rapidly decreased because the electrodes melted and fractured at breakdown. (**b**) $E_{BD}$ as a function of electrode contact force in the type A measurement. All measurements were performed on the same flake with a thickness of 13 nm. (**c**) *IV* curves from the type B measurement under atmospheric environment with a relative humidity of 45 % and a vacuum at 8 ×10$^{-3}$ Pa. The thickness of $h$-BN and the electrode area are ~35 nm and 25 μm$^2$, respectively. (**d**) $E_{BD}$ as a function of electrode area for 35 – 50-nm $h$-BN. An area of 3 × 10$^{-4}$ μm$^2$ is assumed for the type A measurement. The type B and C measurements were conducted under vacuum.

Next, the effect of measurement environment was investigated using the type B method. The surface of $h$-BN is known to be hydrophobic. However, recently, it has been reported that water molecules around carbon nanotubes can be stabilized by forming a layered structure through hydrogen bonding.[23] Therefore, water molecules on $h$-BN should be considered in the examination of dielectric breakdown. **Figure 2(c)** shows the $IV$ curve under atmospheric environment with a relative humidity of 45 % (green) and a vacuum of $8 \times 10^{-3}$ Pa (red). It has been suggested that adsorbed water on $h$-BN can be removed under vacuum.[4] In atmospheric environment, a fluctuating leakage current was measured before breakdown, indicating the existence of a water layer. Lower $E_{BD}$ values often resulted from the dielectric breakdown of $h$-BN assisted by water molecules. Similar behavior was also observed in the type C measurement.

Finally, the effect of the electrode area was investigated using the type B and C measurements. The selection of $h$-BN flakes with the same thickness is quite difficult because one device is fabricated from a single $h$-BN flake. The previous breakdown tests showed no obvious thickness dependence of $E_{BD}$ for $t_{h\text{-BN}} > 20$ nm.[3, 4] Therefore, $E_{BD}$ as a function of electrode area for $h$-BN with $t_{h\text{-BN}} = 35 - 50$ nm is shown in **Figure 2(d)**. The red and blue points indicate type B and C measurements, respectively. Moreover, data for the type A measurement are also plotted using the above mentioned calculated area. Although there is slight variation in the $E_{BD}$ values, the $E_{BD}$ values for the three methods are roughly consistent within experimental error. Therefore, $E_{BD}$ is independent of the electrode area and is ~11 – 12 MV/cm.

Based on the above results, it was found that $E_{BD}$ is not sensitive to contact force or electrode area but strongly depends on the measurement environment, i.e., adsorbed water. Here, let us compare the $E_{BD}$ values obtained by the three methods for various thicknesses of $h$-BN. **Figure 3** clearly shows that $E_{BD}$ gradually decreases with an increase in thickness and seems to saturate to a constant value. Here, although the type A measurement was conducted under atmospheric environment, the $E_{BD}$ of type A agreed well with those of the other two methods. The leakage path from the C-AFM tip to the Pt substrate through the $h$-BN surface, as shown by an arrow in the schematic of the type A measurement in **Figure 1**, is much longer than the thickness of $h$-BN due to the small tip area, unlike in type B and C measurements. Therefore, it is expected that the effect of adsorbed water on $E_{BD}$ is negligible in the type A measurement even for measurements conducted in atmospheric environments.

The reduction in $E_{BD}$ with increased insulator thickness has also been observed for $SiO_2$.[24, 25] The carriers that are accelerated by the electrical field gain energy that is greater in value than that of the band gap, which excites electron-hole pairs and often breaks the bonding. In general, for thicker insulators, the trapped positive charges in the insulator, mainly due to defects, enhance the electrical field near the cathode.[25, 26] This increases the tunneling injection, resulting in breakdown, as schematically shown in the inset of **Figure 3**. Experimentally, the presence of carbon and oxygen impurities in the $h$-BN used in this study was verified in concentrations of less than 10$^{18}$ atoms/cm$^3$.[27] Because the number of defects in the thicker $h$-BN flake is larger than that in the thin $h$-BN flake, assuming the same defect density, more positive charges are trapped in the thicker $h$-BN flake when the same external electrical field is applied. This explains the reduction in $E_{BD}$ with an increase in insulator thickness.

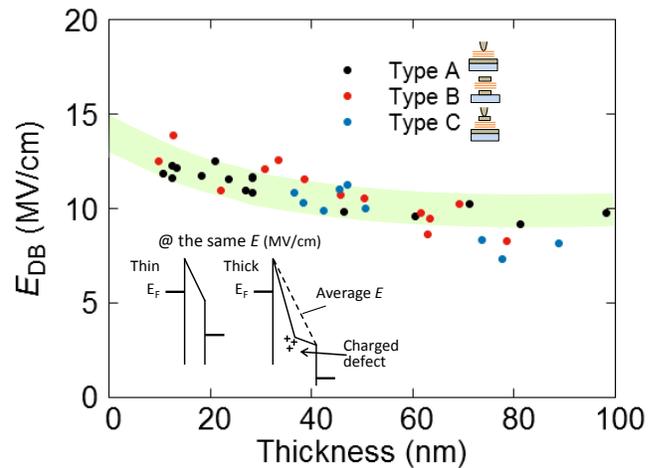

**Figure 3.** $E_{BD}$ as a function of $h$-BN thickness obtained by the three methods. The contact force in the type A measurement is 2 nN. The breakdown tests for type B and C measurements were conducted in vacuum with an electrode area of 9 μm$^2$ for the type B measurement and 130 – 1300 μm$^2$ for the type C measurement.

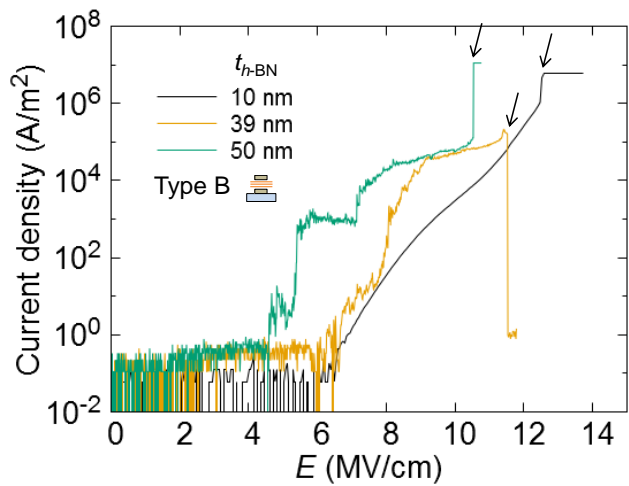

**Figure 4.** Current density as a function of applied electrical field for three different thicknesses of $h$-BN by the type B measurement. The measurements were conducted under vacuum. The electrode area was 9 μm$^2$.

Moreover, the thicker $h$-BN flake seems to have a larger variation in $E_{BD}$ regardless of the electrode structure, especially when the thickness is over 50 nm, as shown in **Figure 3**. The larger variation in $E_{BD}$ can be examined by the *IV* curve before breakdown using the type B measurement. **Figure 4** shows the current density as a function of applied electrical field for $h$-BN flakes with different thicknesses. The FN tunneling current density for 10-nm $h$-BN increases smoothly with an increase in the electrical field, while it fluctuates and suddenly increases before breakdown for thicker $h$-BN flakes. Although the FN tunneling current density should be ideally equal for the three thicknesses, this is not the case because the probability of defect formation is higher for thicker flakes. Conversely, for SiO$_2$, a clear FN tunneling current can be obtained even for thick SiO$_2$ (~60 nm). Therefore, the present fluctuation in FN tunneling current density may originate from defects that already exist in the $h$-BN crystal.[28] In other words, the crystal quality, even for bulk $h$-BN prepared by high-pressure and high-temperature growth,[2] may influence the current fluctuation in FN tunneling. This unstable current seems to cause a larger variation in $E_{BD}$ for thicker $h$-BN, as seen in **Figure 3**.

In summary, a comprehensive evaluation of each electrode structure and environment is indicated in the last row of **Figure 1**. To utilize $E_{BD}$ measurement to compare the film quality of synthesized $h$-BN, the effect of adsorbed water on $h$-BN must be avoided. If adsorbed water is removed, the evaluation of $E_{BD}$ is roughly possible by all three methods. Strictly speaking, the best method is type B because it allows for precise control of the contact area with sufficient current level. Therefore, the crystal quality, as well as $E_{BD}$, may be examined by the current fluctuation in FN tunneling just before breakdown. It is, however, important to keep the interface between the metal and $h$-BN clean. The type A measurement has the advantage of evaluating the local crystalline quality of $h$-BN by examining grain boundaries and wrinkles, as well as single-crystal grains, even in atmospheric environment. The leakage current before breakdown is, however, less than 0.1 pA, which requires the utilization of a high-resolution ammeter. The type C measurement is the easiest to assemble for large-area $h$-BN films, as it does not require the transfer of synthesized $h$-BN if the entire surface area of the catalytic metal substrate is covered by the synthesized $h$-BN.

**SUPPLEMENTARY MATERIAL**

See supplementary material for the optical microscopy images of the whole electrode pattern used in the type B measurement and the tip of the probe used in the type C measurement.

**ACKNOWLEDGMENTS**
This research was partly supported by JSPS Core-to-Core Program, A. Advanced Research Networks, and JSPS KAKENHI Grant Numbers JP25107004, JP16H04343, JP16K14446, & JP26886003.

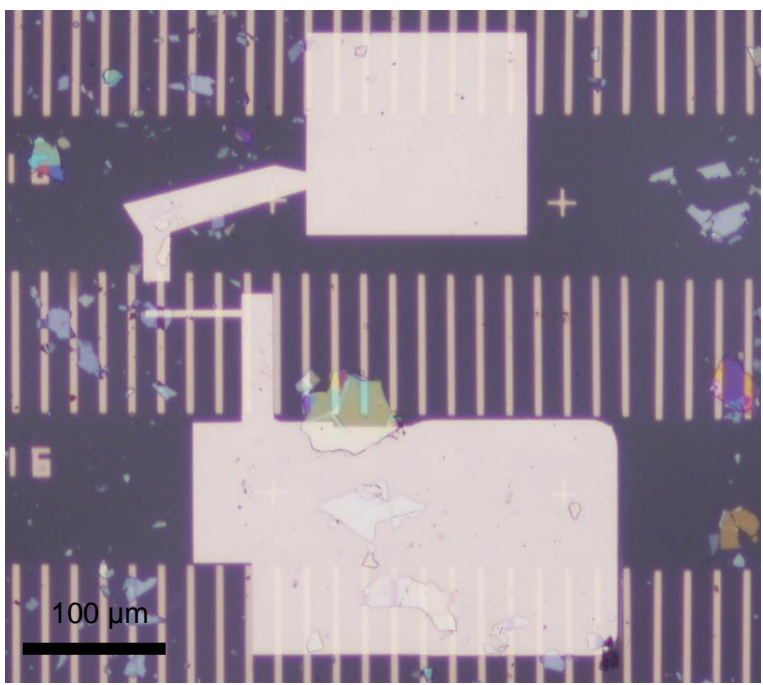

**Figure S1.** Optical microscopy image of the whole electrode pattern used in the type B measurement.

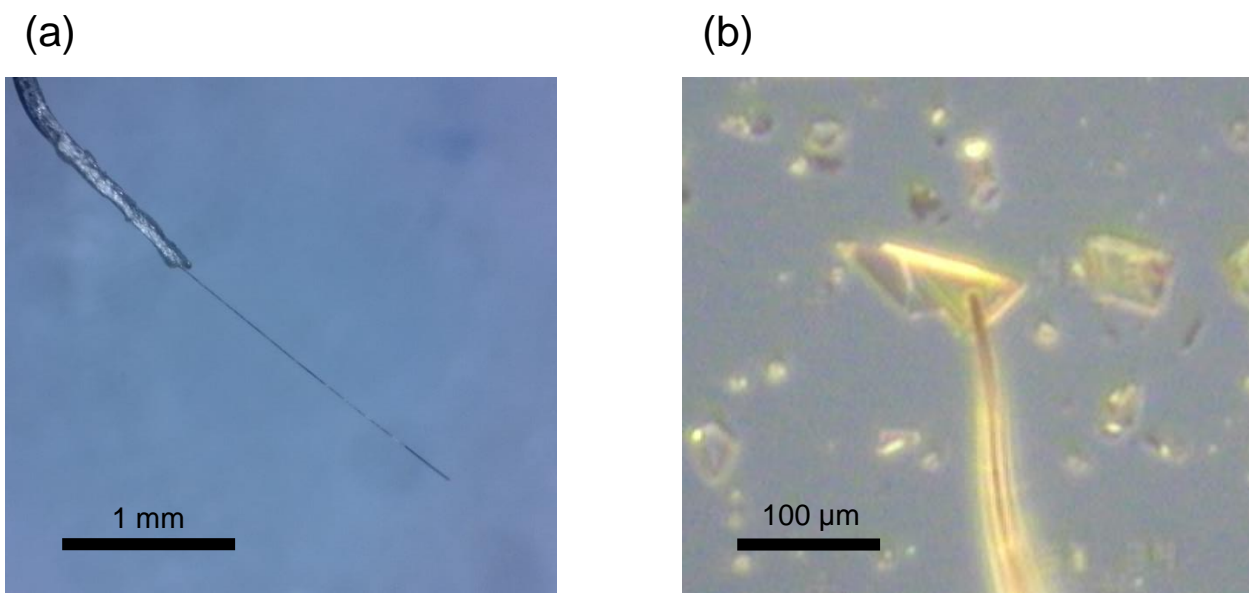

**Figure S2.** Optical microscopy images of the tip of the probe used in the type C measurement (a) and after it approaches the metal electrode on h-BN (b).

1